\documentstyle[epsfig]{aipproc}

\begin{document}

\title{On the Clustering of GRBs on the Sky}

\author{Shiv K. Sethi$^1$, S. G. Bhargavi$^2$, and Jochen Greiner$^3$}
 \address{$^1$Mehta Research Institute, 
Chhatnag Road, Jhusi \\Allahabad 221 506, India \\ 
$^2$Indian Institute for Astrophysics, Koramangla \\
Bangalore 560034, India \\
$^3$Astrophysikalisches Institut Potsdam \\ 
D-14482 Potsdam , An der Sternwarte 16, Germany}

\maketitle

\begin{abstract}
  The two-point correlation of the 4th (current) BATSE catalog (2494
  objects) is calculated. It is shown to be consistent with zero
  at nearly all angular scales of interest.  Assuming
  that GRBs trace the large scale structure in the universe we
  calculate the angular correlation function for the standard CDM
  (sCDM) model. It is shown to be $\le 10^{-4}$ at $\theta \simeq
  5^\circ$ if the BATSE catalog is assumed to be a volume-limited
  sample up to $z \simeq 1$. Combined with the error analysis on the
  BATSE catalog this suggests that nearly $10^5$ GRBs will be needed
  to make a positive detection of the two-point angular correlation
  function at this angular scale.
\end{abstract}

\section*{Introduction}
 
Recent  optical identification of Gamma-ray bursts (GRBs) has
established the cosmological origin of GRBs and redshifts
have  been measured in 9 cases.
(For a comprehensive list of references on this subject see
\cite{greiner}).
However, the physical origin of these bursts, their
environment, and their relationship with other astrophysical objects still
remains an unsolved puzzle. If these bursts are associated with the
underlying large scale structure in the universe, then they should
show clustering in their positions on the sky as expected of cosmological objects.

One way to search for the clustering is to determine the two-point
angular auto-correlation function of the burst positions \cite{blum,hart,lamb}. We compute this quantity for the 4th (current) BATSE catalog (2494
objects) in the next section. In \S 3 we calculate the two-point
correlation function from existing, viable, theoretical models of structure
formation. \S 4 summarizes the main results.

\section*{Two-point correlation function}

Given a two-dimensional distribution of $N$ point objects in a solid angle
$\Omega$, the two-point angular correlation function is defined using
the relation \cite{peebles}:
\begin{equation}
  n_{\scriptscriptstyle \rm DD}= n d\Omega N(1+w(\theta))
  \label{eq:corr}
\end{equation}
Here $n_{\scriptscriptstyle \rm DD} $  is the total number of pairs  between  angular separation
$\theta$ and $\theta + d\theta$;  $n = N/\Omega$; and $d\Omega$ is an
infinitesimal solid angle centered around $\theta$. $w(\theta)$ is the
two-point correlation function. It measures the excess of pairs over
a random Poisson distribution at a given separation
$\theta$. Eq.~(\ref{eq:corr}) is not very convenient for estimating
the two-point correlation function and several alternative estimators of
the two-point angular correlation function have been suggested. We
experimented with several estimators
\cite{peeb1,davis,hamil,landy}. The advantage of using either
\cite{hamil} or \cite{landy} is that  the error on the two-point function is
nearly Poissonian; the leading term in the error for the other two
estimators is $\propto 1/N$, which can dominate over the Poisson
term for large bin size \cite{landy}.  In
this paper  we report results using the estimator given by Landy and Szalay \cite{landy}:
\begin{equation}
 \tilde w(\theta) =  {n_{\scriptscriptstyle \rm DD}
   -2n_{\scriptscriptstyle \rm DR} + n_{\scriptscriptstyle \rm RR}
   \over n_{\scriptscriptstyle \rm RR}}
 \label{eq:corrfun}
\end{equation}
  Here $n_{\scriptscriptstyle \rm DD}$ is the number   of pairs (for a given $\theta$) in the
  GRB  catalog, $n_{\scriptscriptstyle \rm RR}$ is the number of pairs in a mock, random,
  isotropic sample, and  $n_{\scriptscriptstyle \rm DR}$ is the catalog-random
  pair count. 
  The variance  of $\tilde w(\theta)$ is given by:
  \begin{equation}
    \delta \tilde w(\theta)^2 \simeq {1 \over n_{\scriptscriptstyle \rm DD}}. 
    \label{eq:error}
\end{equation}

In Figure~1 (Left Panel) we show the angular correlation function with 1$\sigma$
error bars for current BATSE (2494 objects) catalog. We also plot the 1$\sigma$
 errors (Eq.~(\ref{eq:error})). The main conclusions of our analysis are:

\begin{itemize}
\item[1.] The two-point angular correlation function is consistent
  with zero on nearly all angular scales of interest.
  \item[2.] From Figure~1 (Left Panel) it is seen that at several angular scales a
    1$\sigma$ detection of the correlation function seems to be 
    possible. To make a definitive statement about a detection we need
    to take into account several uncertainties in our analysis. One of
    the  dominant source of uncertainty is the heterogeneity of the sample with
    respect to the error in  angular positions  of the GRBs (the
    localization uncertainty varies from $\simeq
    1^{\circ}\hbox{--}10^\circ$).
    This
    means that errors at $\theta \le 10^\circ$ are much larger than
    seen in Figure~1 (Left Panel). Another
    major source of uncertainty comes from anisotropic exposure
    function of the BATSE instrument, which results in  a non-zero
    correlation function even for a completely isotropic intrinsic
    distribution\footnote{for more details see
      {\tt http://www.batse.msfc.nasa.gov/batse/grb/catalog/}}.
    Though it is
    possible that some of the signal at large angular scales is not
    an artifact, more careful analysis would be required to confirm
    it. 
  \end{itemize}

  \section*{Theoretical Predictions}

  The two-point angular correlation function can be related to the
  two-point three-dimensional correlation function $\xi(r)$ using
  Limber's equation (for details see \cite{peebles}). If we assume that
  the GRBs constitute a volume-limited sample up to  a distance
  $r_{\rm max}$ and that the comoving number density of objects is
  constant, the Limber's equation reduces to:
  \begin{equation}
    w(\theta) = {  \int_0^{r_{max}}
      \int_0^{r_{max}}  r_1^2 r_2^2 dr_1 dr_2  \xi(r_{12},z_1,z_2) \over \left [
        \int_0^{r_{max}}  r^2 dr  \right ] ^2}
    \label{eq:limber1}
\end{equation}
Here
\begin{equation}
  r_{12}^2 = r_1^2 + r_2^2 -2r_1r_2\cos\theta.
  \end{equation}
$r$ is the coordinate distance in an isotropic, homogeneous universe.
The two-point correlation function is related to the power spectrum
$P(k)$ of
the density fluctuations as:
\begin{equation}
  \xi(r,t) = b^2 {1 \over 2 \pi^2}\int_0^\infty k^2 dk P(k,t) {\sin(kr)
    \over kr}. 
  \label{eq:powspec}
\end{equation}
$b$, the bias factor, denotes the clustering of visible matter
relative to the dark matter. While its absolute value is still
uncertain, the relative bias between nearby  rich clusters of galaxies and
optically-identified galaxies is $\simeq 5$. And hence if GRBs
originate in clusters rather than  ordinary galaxies their
correlation can be 25 times larger.
In this paper, we use the linear perturbation theory predictions for 
$P(k,t)$. We have checked that for the angular scales of interest
($\theta \ge 5^\circ$) it is a reasonable assumption. We use the
BBKS fit \cite{bardeen}  for   the
linear power spectrum of the  standard CDM (sCDM) model and some of
its variants. We normalize the power spectrum requiring $\sigma_8 =
0.7$. The time dependence of linear power spectrum is $P(k,t) \propto
(1+z)^{-2}$, which is also the time dependence of the two-point
correlation function. It should be noted that in general the two-point
correlation function depends on both the separation between two points
and their redshifts, as indicated in Eq.~(\ref{eq:limber1}). However, 
the two-point correlation function is
negligible for points separated by a large enough redshift
difference. Therefore, for most purposes $\xi(r,t) \propto
(1+z)^{-2}$, where $z$ refers the redshift of any of the two points.

In Figure~1 (Right Panel) we show the theoretically predicted angular two-point
correlation function for sCDM model. The bias $b$ is taken to be one.
If observed GRBs constitute a complete sample up to $z \simeq 1$
and they are assumed to be associated with highly biased structures
like rich clusters, the value of correlation function is $\le 10^{-4}$
at $5^\circ$. This is the smallest angular scale at which information is
possible in the BATSE catalog. At larger angles the correlation function
typically scales as $\theta^{-1}$.

\begin{figure}[b!] 
\centerline{\epsfig{file=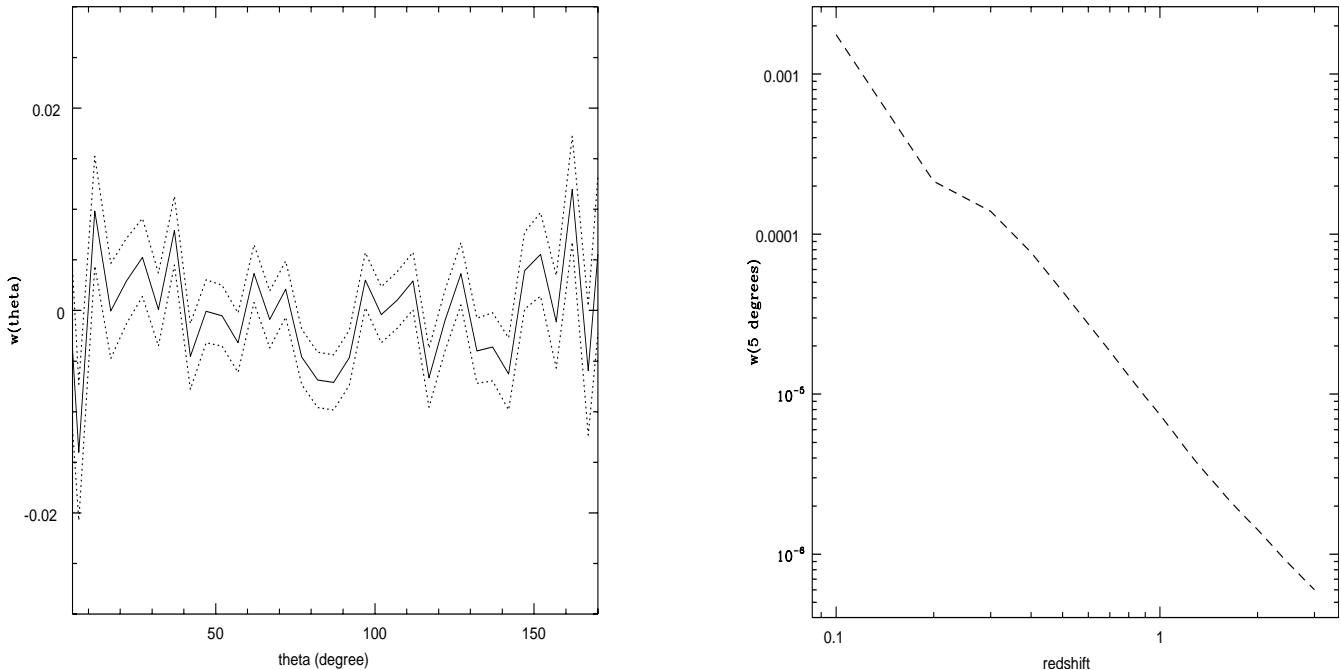,height=3.5in,width=7in}}
\vspace{10pt}
\caption{{\it Left Panel\/}: The two-point angular correlation function for the BATSE
  catalog (2494 objects) and  the 1$\sigma$ error bars  are shown. The
  {\it solid \/} line corresponds to the two-point correlation function. The {\it
    dotted \/} lines show the 1$\sigma$ errors given by Eq.~(\ref{eq:error}). 
{\it Right Panel \/}: Theoretical prediction for the two-point angular correlation
  function is shown for the sCDM model as a function of depth
  (redshift) of the sample. The quantity plotted is the absolute value
  of the two-point correlation function at $\theta = 5^\circ$. }
\label{fig1}
\end{figure}

\section*{Conclusions and Summary}

The two-point correlation function of the 4th (current) BATSE catalog (2494
objects) is consistent with zero at nearly all angular scales that can be
probed in the BATSE catalog.
This result is consistent with theory if the GRBs are assumed to
trace the dark matter distribution with some bias and are a complete
sample up to $z \simeq 1$.

When can a detection of the two-point correlation function become
possible? The error in the two-point
correlation function scales as $n_{\scriptscriptstyle \rm DD}^{-1/2}$
(Eq.~\ref{eq:error}) and  $n_{\scriptscriptstyle \rm DD} \propto N^2$, $N$ being the number of
objects in the catalog. Therefore the error in estimating the two-point correlation
function scales as $1/N$. Theory suggests that the value of correlation function at
$\theta = 5^\circ$ is $\le 10^{-4}$ if the GRB sample is assumed to be
complete up to $z \simeq 1$. We check that $n_{\scriptscriptstyle \rm DD}$ at $\theta \simeq 5^\circ$
is $\simeq 10^{-2}$ times the total number of pairs ($\simeq N^2/2$) in
the  GRB sample. This would suggest that a detection might
become possible at this angular scale when the number of objects in
the sample exceeds $10^5$.

Future surveys like HETE-II and SWIFT will localize the GRBs to a few
arc-minutes. This means smaller angular scales could be probed. And as
the theoretically-predicted two-point correlation function scales as $\sim \theta^{-1}$, the
probability of detection will increase. SWIFT will detect nearly 1000
objects over a period of 3 years with an angular resolution $\le
1''$. However, though the two-point correlation function is large at
these angular scales, the average separation between 1000 objects on
the complete sky is $\simeq 6^\circ$. Therefore as long as $w(\theta)
\le 1$, the probability of finding an object within a few arcseconds
of the other is negligible. It is possible that $w(\theta) \gg
1$ at sub-arcsecond scales. However, detailed analysis, taking into account the
non-linear correction to the power spectrum of density perturbation,
is needed to make  precise theoretical predictions  for the future surveys.

\bigskip
\noindent {\it Acknowledgements:
JG is supported by the German Bun\-des\-mi\-ni\-sterium f\"ur Bildung,
Wissenschaft, Forschung und Technologie (BMBF/DLR) under contract
No. 50 QQ 9602 3. RJG acknowledges a travel grant from DFG (KON 1973/1999 
and GR 1350/7-1) to attend this conference. SKS thanks S. Bharadwaj
for pointing out  an error. We  thank D. H. Hartman  for valuable
comments on the anisotropic exposure function of BATSE.}

\end{document}